\documentstyle[multicol,epsf,prl,aps]{revtex}
\tighten
 
\begin{document}
\title{Shot Noise through a Quantum Dot in the Kondo Regime}

\author{Yigal Meir$^{1,2}$ and Anatoly Golub$^1$}
\address{$^1$ Physics Department, Ben Gurion University, Beer Sheva 84105,
Israel
\\$^2$ The Ilse Katz Center for Meso- and Nanoscale Science and Technology,\\
 Ben-Gurion University, Beer Sheva 84105, ISRAEL}
\date{\today}
\maketitle
\begin{abstract}
The shot noise in the current through a quantum dot is calculated
as a function of voltage from the high-voltage, Coulomb blockaded
regime to the low-voltage, Kondo regime. Using several
complementary approaches, it is shown that the zero-frequency shot
noise (scaled by the voltage) exhibits a non-monotonic dependence
on voltage, with a peak around the Kondo temperature. Beyond
giving a good estimate of the Kondo temperature, it is shown that
the shot noise yields additional information on the effects of
electronic correlations on the local density of states in the
Kondo regime, unaccessible in traditional transport measurements.

\end{abstract}
\pacs{Pacs numbers: 73.23.Hk, 72.15.Qm, 73.23.Hk}

% definitions
\def\squote{}
\def\quote#1#2#3#4{\squote {#1,\ {\sl#2}\ {\bf#3}, #4}.\par}
\def\qquote#1#2#3#4{\squote {#1,\ {\sl#2}\ {\bf#3}, #4};}
\def\nquote#1#2#3#4{\squote {#1,\ {\sl#2}\ {\bf#3}, #4}}
\def\book#1#2#3{\squote { #1,\ in {\sl#2}, edited by #3}.\par}
\def\nbook#1#2#3{\squote { #1,\ in {\sl#2}, edited by #3}}
\def\bbook#1#2#3{\squote { #1,\ in {\sl#2}, edited by #3}.}
\def\trans#1#2#3{[ {\sl #1} {\bf #2},\ #3 ]}
\def\prl{{\sl Phys. Rev. Lett.}\ }
\def\PRL{{\sl Phys. Rev. Lett.}\ }
\def\apl{{\sl Appl. Phys. Lett.}\ }
\def\APL{{\sl Appl. Phys. Lett.}\ }
\def\euro{{\sl Europhys. Lett.}\ }
\def\ds{\displaystyle}
\def\pr {{\sl Phys. Rev.}\ }
\def\ksi{\xi}
\def\l{\ell}
\def\sss{\scriptscriptstyle}
\def\w{\omega}
\def\e{\epsilon}
\def\deriv{\partial}
\def\T{{\cal{T}}}
\def\fl {f_{\sss L}(\e)}
\def\fr {f_{\sss R}(\e)}
\def\flw {f_{\sss L}(\w)}
\def\frw {f_{\sss R}(\w)}
\def\flr {f_{\sss L(R)}}
\def\mul {\mu_{\sss L}}
\def\mur {\mu_{\sss R}}
\def\nl {N_{\sss L}}
\def\nr {N_{\sss R}}
\def\mulr {\mu_{\sss L/R}}
\def\sp                                {
     \hskip 0.02in                            }
\def\ss                               {
     \scriptstyle                       }
\def\sss                               {
     \scriptscriptstyle                       }
\def\raise                             {
     {\sss +}                            }
\def\k                               {
     k                            }
\def\w                               {
     \omega                             }
\def\c                               {
     {\bf c}                            }
\def\spin                            {
     \sigma                             }
\def\spinbar                         {
     {\bar \spin}                  }
\def\cspin                               {
     {\bf d}_{\spin}                   }
\def\cspindag                          {
     {\bf d}_{\spin}^{\raise}               }
\def\fspin                               {
     {\bf f}_{\spin}                   }
\def\fspindag                          {
     {\bf f}_{\spin}^{\raise}               }
\def\fspindagbar                          {
     {\bf f}_{\spinbar}^{\raise}               }
\def\b                                   {
     {\bf b}                   }
\def\bdag                                   {
     {\bf b}^{\!\raise}                   }
\def\btwodag                                   {
     {\bf b}_2^{\!\raise}                   }
\def\fspindag                          {
     {\bf f}_{\spin}^{\raise}               }
\def\cbar                                {
     {\bf c}_{\spinbar}                   }
\def\cbardag                          {
     {\bf c}_{\spinbar}^{\raise}               }
\def\ckspin                               {
     {\bf c}_{\k\spin}                   }
\def\ckspindag                            {
     {\bf c}_{\k\spin}^{\raise}              }
\def\Vkspin                               {
     V_{\k\spin}                   }
\def\Vkbar                            {
     V_{\k\spinbar}                   }
\def\espin                           {
     \epsilon_{\spin}                   }
\def\ebar                        {
     \epsilon_{\spinbar}                }
\def\ekspin                           {
     \epsilon_{\k\spin}                   }
\def\ekbar                        {
     \epsilon_{\k\spinbar}                }
\def\sumk                            {
     \sum_{\k{\sss \in L,R} }                     }
\def\sumspin                         {
     \sum_{\spin}               }
\def\sumkspin                         {
     \sum_{\spin;\k{\sss \in L,R} }               }
\def\sumlr                         {
     \sum_{\sss L,R }               }
\def\sumklorr                         {
     \sum_{\k {\sss \in L(R)} }               }
\def\nup                              {
     n_{\sss\uparrow}                         }
\def\ndown                            {
     n_{\sss\downarrow}                         }
\def\nbar                             {
     \langle n_{\spinbar} \rangle              }
\def\nspin                            {
     \langle n_{\spin} \rangle              }
\def\nspinprime                            {
     \langle n_{\spin'} \rangle              }
\def\Gt                               {
     G_{\spin}(t)                         }
\def\Grt {G_{\spin}^{\,r}(t)}
\def\Gkt                               {
     G_{\k\spin}(t)                         }
\def\Gl        {
     G_{\spin}^<                         }
\def\Gg        {
     G_{\spin}^>                         }
\def\Glg       {
     G_{\spin}^{\stackrel{\sss <}{\sss >}}  }
\def\Gtwo                              {
     G_{\spinbar \spin}(t)                         }
\def\Ge                               {
     G_{\spin}(\e)                         }
\def\Grw                               {
     G^{\,r}_{\spin}(\w)                         }
\def\Glessw        {
     G_{\spin}^<(\w)                         }
\def\Fw                               {
     F_{\spin}(\w)                         }
\def\fbar   {\bar f_{\spin}(\w)}
\def\sigzero                          {
     \Sigma_{{\sss 0}\spin}              }
\def\sigone                          {
     \Sigma_{{\sss 1}\spin}              }
\def\cond                             {
     \sigma                               }
\def\Gamalr                            {
     \Gamma^{\sss L(R)}_{\spin}               }
\def\Gamazl                            {
     \Gamma^{\sss L}_{{\sss 0}\spin}              }
\def\Gamazr                            {
     \Gamma^{\sss R}_{{\sss 0}\spin}              }
\def\Gamazlr                            {
     \Gamma^{\sss L(R)}_{{\sss 0}\spin}              }
\def\Gama                            {
     \Gamma_{\spin}               }
\def\Gamal                            {
     \Gamma^{\sss L}_{\spin}               }
\def\Gamar                            {
     \Gamma^{\sss R}_{\spin}               }
\def\Gamblr                            {
     \Gamma^{\sss L(R)}_{{\sss 1}\spin}               }
\def\Gambl                            {
     \Gamma^{\sss L}_{{\sss 1}\spin}               }
\def\Gambr                            {
     \Gamma^{\sss R}_{{\sss 1}\spin}               }

\begin{multicols}{2}
The Kondo effect has become one of the main paradigms of condensed
matter physics as
 it is one of the simplest models that exhibit many-body correlations\cite{hewson}.
 The original model was devised to explain the non-monotonic
 resistivity of metal due to  enhanced scattering by magnetic impurities
 below the Kondo temperature. This effect, however, was also predicted to play
 a dramatic role in transport through quantum dots
 \cite{glazman88,hershfield91,ournca}, due to the enhancement
 of the local density of states at the Fermi energy.
Indeed, the recent observation of the Kondo effect in transport
through a quantum dot \cite{goldhaber,reviews} has paved the way
for a new class of experimental investigations of strongly
correlated electrons in general and the Kondo effect in
particular. These and later experiments
\cite{goldhaberNRG,unitary,phase,irradiation,even,pustil}
demonstrated the ability to exploit the tunable physical
characteristics of the  quantum dot in order to yield important
information on Kondo systems, information unavailable from
experiments in bulk systems. Such studies,  for example, included
 the full crossover between the Kondo limit, the mixed valence
regime and the non-Kondo limit
% with a critical comparison to
% numerical renormalization group calculations
 \cite{goldhaberNRG},
the emergence of the unitarity limit \cite{unitary},  the
determination of the phase of the transmission coefficient through
such an Anderson impurity\cite{phase}, the study of the Kondo
effect under external irradiation \cite{irradiation}, and even
surprises such as
 the observation of the Kondo effect for
an integer-spin dot \cite{even} and the enhancement of the Kondo
effect by a magnetic field\cite{pustil}. These new probes enhanced
our understanding of a
 the Kondo system and provided critical tests of various theoretical
 approximations,  an imperative step towards better understanding of
strongly correlated electron systems. Nevertheless, detailed
experimental information on how electronic correlations affect the
density of states is still lacking.

In this letter we propose and explore theoretically another
experimental tool to probe the Kondo regime - shot noise
measurements. Such measurements proved very successful in
mesoscopic structures formed in other strongly correlated electron
systems,  such as a fractional Hall liquid \cite{glattli} or a
superconductor\cite{superconductor}. Here we demonstrate that the
noise measurements yield  additional information on the structure
of the local density of states, information unavailable by usual
transport measurements.  In addition shot noise yields a direct
estimate of the Kondo temperature.

 Current noise, defined as
\begin{equation}
S(t)\equiv<I(t)I(0)>-<I>^2, \label{S}
\end{equation}
or  alternatively,
 its Fourier transform, $S(\w)$,   has been studied
extensively in the context of mesoscopic systems in the last
couple of
 decades  \cite{buttiker} ($I$ above is the current operator).
While the equilibrium zero-frequency noise $S_0\equiv S(\w=0)$ can
be related to the conductance via the fluctuation-dissipation
theorem, and does not carry additional information,  the
zero-frequency noise out of equilibrium (shot noise) can yield
information on charge fluctuations in the mesoscopic system. Since
for voltage bias much larger than temperature,  finite temperature
can be ignored,
  we concentrate in the following on $T=0$,  where the
thermal noise vanishes and the only contribution to the current
noise is shot noise. Due to the lack of a single accurate method
that can describe the Kondo system out of equilibrium in all
regimes, we employed five different methods that span all
physically relevant regimes.

The current $I$ through a quantum dot in the presence of a DC voltage bias
$V$ can be directly related to
 the transmission coefficient through the dot, $\T(\e,V)$, which, in turn,
is proportional to the local density of states $\rho(\e,V)$
\cite{landauer,drop}, $I = {2e/h}  \int_0^{eV} \T(\e,V) d\e$,
where the factor $2$ is due to spin-degeneracy. For noninteracting
electrons, $\T(\e,V)$ is voltage independent, and thus the
differential conductance, $dI/dV$, yields directly $T(\e=V)$,
 and hence the full local density of states.
In the present case, however, the Kondo peak at the Fermi energy
is dramatically affected by voltage \cite{hershfield91,ournca}.
Thus, while its structure is of major importance, current
measurements
  cannot yield the energy-dependent
transmission coefficient $\T(\e,V)$, but rather its average over
 a scale $0\le \e\le eV$. Since $\rho(\e,V)$ is of major interest,
due to the Kondo resonance, an experimental probe of, e.g., its
higher moments is highly desirable.

For noninteracting electrons the shot noise can be expressed in
terms of the transmission coefficients\cite{lesovik},
\begin{equation}
S_0(V) = {{4 e^2 |V|}\over{h}} \int_0^{eV} {\T(\e)}
\left[1-{\T(\e)}\right] d\e . \label{noise}
\end{equation}
One might expect anomalous shot noise dependence on voltage or
temperature due to the Kondo effect because of the following
argument. In the absence of the Kondo effect (e.g. when
temperature or bias are much larger than the Kondo temperature)
the conductance through a quantum dot,  or the effective
transmission coefficient,  is suppressed due to the Coulomb
blockade (except at the Coulomb blockade peaks),
  and $\T(1-\T)$ is very small. With
lowering of temperature or voltage,  the conductance is enhanced
(for an odd number of electrons on the dot),  leading to an
increase in the shot noise. However,  at  the unitarity limit,  at
zero temperature and linear response, $\T=1$ (for the special case
of symmetric barriers),  and the noise again vanishes. Thus one
may expect a non-monotonic dependence of the noise (scaled by $V$)
on voltage, for example. As we will show below,  such
non-monotonicity should  indeed be observed.

Shot noise through a quantum dot has been studied in the past,
with an emphasis on the Coulomb blockade regime
\cite{buttiker,hershfield}. These studies revealed indeed that the
shot noise is quite small at the conductance valleys at high
voltages (or temperatures). Several attempts to look at the Kondo
regime has been made.  Perturbative results were reported
\cite{yamaguchi}, while exact solution is available for a
particular limiting case (not quite relevant to the present case)
\cite{schiller}. Here we report calculations made in the
noncrossing approximation (NCA) \cite{bickers},  augmented by
high-voltage perturbation theory, renormalization group (RG)
calculations,  perturbation theory around the zero-voltage,
Fermi-liquid point, and by slave-boson mean field theory (SBMFT).
While NCA is valid for a wide range of voltages, including $eV,T<
T_K$ (except a small region for small $eV$ and $T$),  the RG
calculation is valid at high voltages, while the SBMFT is valid at
small voltages. Thus our methods complement each other and give a
consistent view of the dependence of shot noise on voltage.

Our starting point is the infinite-$U$ single-impurity  Anderson
Hamiltonian,
\begin{eqnarray}
 H = \sumkspin\!\! \ekspin \ckspindag \ckspin
  + \sumspin \e_0 \cspindag \cspin \nonumber\\
+ \sumkspin \!
       (\Vkspin \ckspindag \cspin + h.c.),
\label{eq:H}
\end{eqnarray}
where $\ckspindag (\ckspin)$  creates (destroys) an electron with
momentum $k$ and spin $\spin$ in one of the two leads, and
$\cspindag (\cspin)$  creates (destroys) a spin-$\spin$ electron
on the quantum dot. Coulomb interactions among electrons, in the
limit of $U \rightarrow \infty$, forbid double occupancy of the
quantum dot. The last term describes the hopping between the leads
and the dot, and determines this coupling $\Gamma=\Gamal+\Gamar$,
via $\Gamalr(\w)  = 2\pi\! \sumklorr |\Vkspin|^2
   \delta(\w - \ekspin)$.

We start by studying the limiting cases, $eV>>T_K$ and $eV<<T_K$.
In these regimes it is more convenient to study the Kondo
Hamiltonian, obtained from the Anderson Hamiltonian (\ref{eq:H})
by a Schrieffer-Wolf transformation. For convenience  one performs
first a unitary transformation where the dependence on external
voltages  is shifted to the couplings $\Gamalr$\cite{glazman}. The
resulting Kondo Hamiltonian is
\begin{equation}
H_K=\sum_{k\sigma\\ j=L,R}
\xi_{k}\psi^{\dagger}_{kj\sigma}\psi_{kj\sigma} + \sum_{j,j'{\sss
= L,R}}J_{jj'}(t)\psi_{j}^{\dagger}(0)\hat{P}\psi_{j'}(0),
\end{equation}
with $\hat{P}=\frac{1}{4}\hat{I}+\vec{S}\cdot\vec{s}$ and where
$\hat{I}$, $\vec{s}$ and $\vec{S}$ are identity, electron spin
operator of the leads and and of the electron on the impurity,
respectively. The coupling parameters $J_{jj'}$ are related to
those  of the Anderson Hamiltonian and due to the unitary
transformation depend on time
\begin{equation}
J_{LR}=J_{RL}^{*}=J_{0}\exp[\frac{iteV}{\hbar}],\,\,\,
J_{0}=\frac{\sqrt{\Gamma_{L}\Gamma_{R}}}{\pi\nu\e_0},\,\,\,\,
J_{jj}=\frac{\Gamma_{j}}{\pi\nu\e_0},
\end{equation}

where $\nu$ is the density od states in the leads. Using the
Keldysh formalism \cite{kamenev} we can evaluate the noise $S_0$
to third order in the coupling,
\begin{eqnarray}
S_0(V)&=& \frac{2e^3|V| }{2h}(\pi\nu J_{0})^{2}\nonumber \\ &+&
\frac{3e^3|V|}{h}(\pi\nu J_{0})^{2} [1+2\nu
(J_{LL}+J_{RR})\log\frac{D}{eV}] \label{fixedpoint}
\end{eqnarray}
Here D is the effective bandwidth, and the leading term is
separated into two terms, expecting the RG procedure. Indeed, to
get an expression valid also for lower voltages, one can use the
RG to sum up the diverging logarithms (second line in
(\ref{fixedpoint})), leading to
\begin{eqnarray}
S_0(V) = \frac{3 e^3\gamma |V|}{4h}\left[\frac{\pi} {
\log(eV/T_{K})}\right]^2
\end{eqnarray}
with $\gamma\equiv 4\Gamma_L \Gamma_R /(\Gamma_L+\Gamma_R)^2$, and
where the relation between the renormalized coupling $J_{0}(V)$
and $T_K$,
 $J_{0}(V)=\sqrt{\gamma}/2\nu
\log(eV/T_{K})$  \cite{glazman} was used.

In the other limit, of small voltage, $eV<<T_K$, one can expand
around the strong coupling Fermi-liquid fixed point
\cite{nosieres}. Identifying the current operator around this
point \cite{glazman}, a straightforward, lengthy calculation
yields
\begin{equation}
S_0(V)=\frac{2e^3}{h}|V|\frac{(\Gamma_{L}-\Gamma_{R})^2}
{(\Gamma_{L}+\Gamma_{R})^2} + \frac{4e^3
\gamma}{3h}|V|(\frac{eV}{T_{K}})^2.
\end{equation}
In the case of symmetric barriers $\Gamal=\Gamar$, where the
effective transmission coefficient is unity, we find that
$S_0/V\rightarrow0$ as $V\rightarrow0$, in agreement with the
expectation from the noninteracting noise formula.

To expand the treatment beyond second order in voltage, we
transform the Anderson Hamiltonian (\ref{eq:H}) into a new
Hamiltonian, expressed in terms of new local operators
\cite{barnes}.
 These operators create the three
possible states of the site: a boson operator $\bdag$, which
creates an empty site, and two fermion operators, $\fspindag$,
which create the singly occupied states. The ordinary electron
operators on the site, which transform the empty site into a
singly occupied site or vice versa, are decomposed into a boson
operator and a fermion operator, $ \cspin(t) = \bdag(t)\,\fspin(t)
$. The additional constraint, that the number of fermions and
bosons is equal to one, prevents double occupancy on the site, as
required by the $U\rightarrow\infty$ limit.

In the slave-boson representation, the Hamiltonian for the
infinite-$U$ Anderson model becomes
\begin{eqnarray}
 H = \sumkspin\!\! \ekspin \ckspindag \ckspin
  + \sumspin \espin \fspindag \fspin \nonumber\\
+ \sumkspin \!
       (\Vkspin \ckspindag \bdag  \fspin + h.c.),\label{eq:Hboson}
\end{eqnarray}
where the Hamiltonian only operates in the subspace where the
total number of fermions and bosons is one. The advantage of the
slave-boson representation is that the hopping term, which is
usually the smallest physical term, can be treated perturbatively
using standard diagrammatic techniques.

We first apply  the SBMFT, a theory that is known to give the
correct qualitative behavior at low temperatures and voltages
($eV,T\leq T_K$). In this theory, motivated by a large-$N$
expansion, where $N$ is the degeneracy of the level, one replaces
the boson operator by it classical, non-fluctuating value, giving
rise to an effective resonant-tunneling model, whose parameters
are obtained self-consistently \cite{barnes}. These equation were
numerically solved for a set of parameters leading to the same
Kondo temperature used for the above calculations, and indeed
reduced to the results of the Fermi-liquid perturbation theory for
small voltages.

To bridge the small-V treatment with the large-V one,  we next
employ the NCA, which has been used successfully to treat the
infinite-$U$ Anderson model in \cite{bickers} and out
\cite{ournca} of equilibrium. At lowest order in perturbation
theory the boson self-energy involves the fermion propagator while
the fermion self-energy involves the boson propagator. By using
the two relations self-consistently, one obtains a set of coupled
integral equations, which can be solved numerically. Solving these
self-consistent equations corresponds to summing a subset of
diagrams to all orders in the hopping matrix element. It can be
shown\cite{bickers} that all diagrams of leading order in $1/N$,
where $N$ is the number of spin degrees of freedom, are included
in this subset. Therefore, the non-crossing approximation is
expected to be a quantitative approach in the limit of large $N$.
For the case $N=2$, of interest for quantum dots,
Cox\cite{bickers} has shown that the calculated equilibrium
occupancy and susceptibility agree with the exact Bethe ansatz
results to within the 0.5\% convergence accuracy of the NCA. For
the electrical current calculations \cite{ournca},
 which use an extension of the above equations to include Keldysh
Green functions \cite{kamenev}, at worse an overestimate of $15\%$
 on the linear response conductance has been observed\cite{ournca}.

Here we employ a two-step approximation. The general noise diagram
involves two single-particle Green functions, which, in the first
step, are decoupled, i.e. vertex diagrams are neglected.  In the
second step, these single-particle Green functions are replaced by
their NCA values \cite{ournca}. This procedure ensures that our
expression for the noise obeys exactly  the zero-voltage
fluctuation-dissipation theorem. All the calculations were done
for $T=10^{-4}\Gamma$, well below the Kondo temperature.

\vskip  -1.0 truecm
\begin{center}
\leavevmode \epsfxsize=3.9in
%\epsfbox{dft_fig.ps}
\epsfbox[62 -18 626 452]{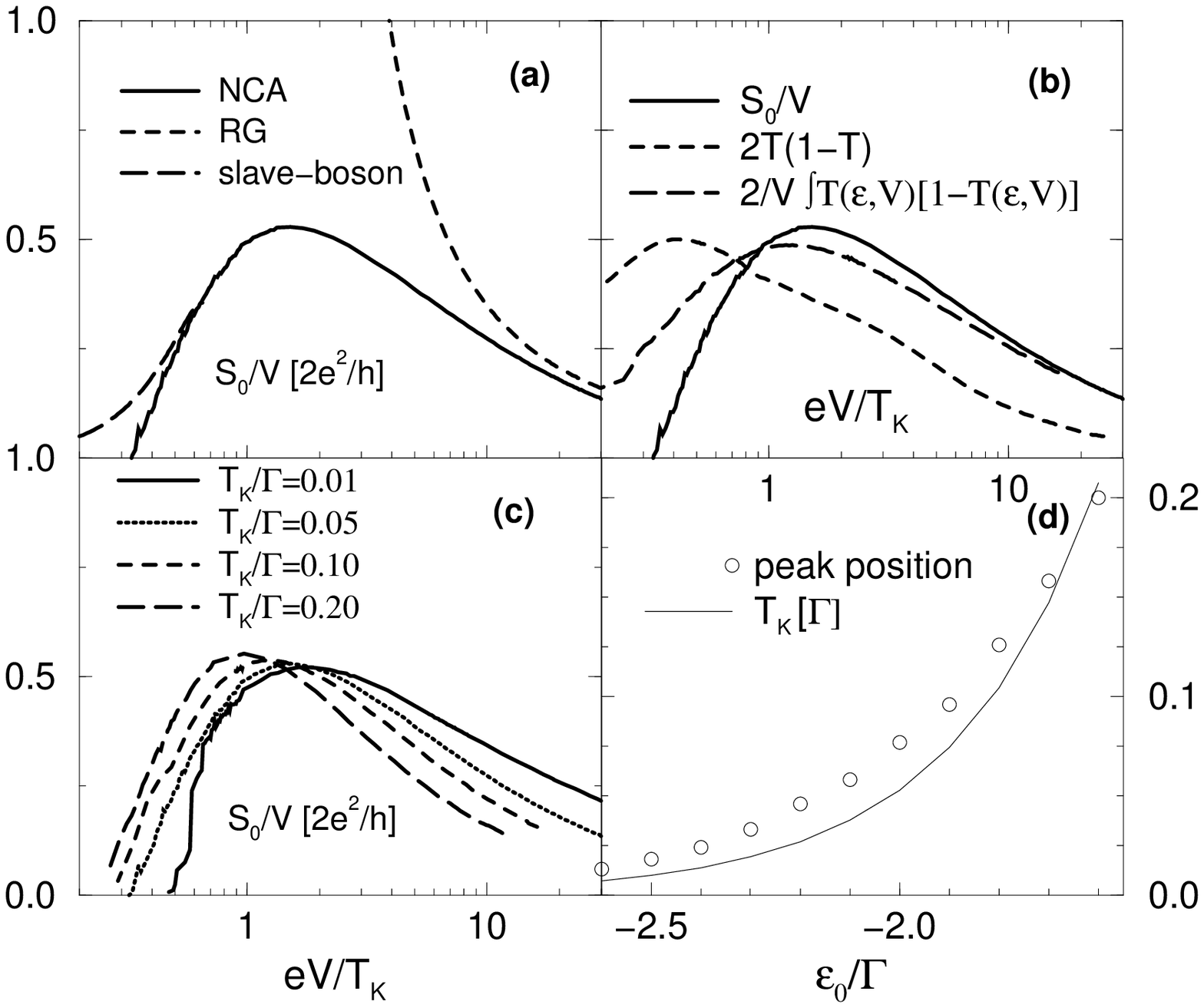}
\end{center}
\begin{small}
\vskip -1.0 truecm Fig. 1. The main results of this work (see
text). (a) Three of the methods used to estimate the noise in the
different regimes. (b) Comparison of the results of the NCA
approximation to the naive noninteracting formula Eq.(\ref{noise})
without and with voltage-dependent transmission coefficient,
obtained from the differential conductance. (c) $S_0/V$ for
different level energies (and different Kondo temperatures). (d)
Comparison of the maxima of $S_0/V$ to the Kondo temperature.
\end{small}
\vskip 0.5 truecm

Figure~1 summarizes our main findings,  depicting $S_0/V$  as a
function of $\log(eV/T_K)$,  the natural parameter in the Kondo
regime. In (a)  we
 plot results from three different approaches. The
 RG calculation agrees with the NCA result at large
 voltages,  but becomes unphysical once $eV$ is about $10T_K$,  while
 the SBMFT,  which agrees with perturbation
 theory around the Kondo fixed point at low voltages,  complements
 our NCA calculations for $eV\ll T_k$,  where the NCA approximation
 stops to be valid. Thus we have a quantitative picture of the noise on
 the whole range of voltages. Here and in (b) the NCA results are for
 $\e_0=-2\Gamma$, resulting
 in $T_K\simeq 0.005\Gamma$. Fig.1(b) demonstrates that $S_0/V$
 exhibits a maximum of a value about $1/2$, as one might expect from the
 heuristic arguments above. Indeed one can notice the similarity of $S_0/V$ to
 Eq.(\ref{noise}) (long-dashed line), where $T(\e,V)$ was calculated from the
  NCA for the same bias. Thus the noise gives a direct measure of
  the second moment of the density of states. In order to check
  how sensitive the noise is to the effects of electronic correlations,
  we also plot in the same figure (short-dashed line) the expected noise for
  a noninteracting system, with the same differential conductance.
  The significant difference between the two curves is due to
  effect that because of the interactions, the  density of states is strongly dependent
  on  the bias
 voltage. This also demonstrates that the information available
 in noise measurements cannot be obtained in usual transport
 experiments.
 In (c) we plot $S_0/V$ for different values
 of $\e_0$, the electron level energy,  leading to values of $T_K$ differing by a factor
 of 20. Nevertheless,  the value of $V/T_K$ where $S_0/V$ is maximal changes
 by less than a factor of 2,  demonstrating that the peak position gives a reliable
 estimate of  the value of the Kondo temperature. This point is further demonstrated
 in (d), where we plot the dependence of the position of the peak on voltage and
 compare to the known NCA values of $T_K= (\Gamma D^2/2\pi\e_0)^{1/2}\exp\left[-\pi\e_0/\Gamma\right]$.
   Thus shot noise
 measurements lead to a straightforward determination of the all
  important Kondo temperature,
 which in usual current measurements can only be determined by
 further nontrivial analysis.

To conclude, we have used several methods which give a consistent
determination of the current noise through a quantum dot in the
Kondo regime, demonstrating the importance of the electronic
correlations. It has been shown that the noise yields additional
information about the Kondo state, and we hope that this work will
indeed motivate experimental effort in this direction.

We acknowledge support from BMBF. We thank N. S. Wingreen for
providing us with the  the NCA program.

\end{multicols}

\begin{references}
\bibitem{drop}
These expressions assume that that the change in chemical
potential due to the voltage drop takes place in one lead. If the
voltage drops symmetrically over the two leads, then the integral
runs from $-V/2$ to $V/2$.
\end{references}
\end{document}